Title Page

# Plasmon-driven Ultrafast and Highly Efficient Saturable Absorption for Ultrashort Pulse Generation Based on 2D $V_2C$

Yingwei Wang[1,*], Li Zhou, Quan Long[1], Xin Li[1,4], Haolin Chang[1], Ning Li[3], Yiduo Wang[1,2,*], Bei Zhang[4], Zhihui Chen[1], Zhongjian Yang[1], Si Xiao[1], Chujun Zhao[3,*], Shuangchun Wen[3], Jun He[1,*]

[1]Hunan Key Laboratory of Nanophotonics and Devices, School of Physics, Central South University, Changsha 410083, China

[2]Beijing National Laboratory for Condensed Matter Physics, Institute of Physics, Chinese Academy of Sciences, Beijing 100190, China

[3]Key Laboratory for Micro- and Nano-Optoelectronic Devices of Ministry of Education, College of Physics and Microelectronic Science, Hunan University, Changsha 410082, China

[4]School of Physics Science and Technology, Xinjiang University, Urumqi, Xinjiang 830046, China

[*] Email: wyw1988@csu.edu.cn; yiduowang@iphy.ac.cn; cjzhao@hnu.edu.cn; junhe@csu.edu.cn;

**Abstract**

Plasmon-driven ultrafast nonlinearities hold promise for advanced photonics but remain challenging to harness in two-dimensional materials at telecommunication wavelengths. Here, we demonstrate few-layer $V_2C$ MXene as a high-performance saturable absorber by leveraging its tailored surface plasmon resonance. Combining transient absorption spectroscopy and first-principles calculations, we unveil a plasmon-driven relaxation mechanism dominated by interfacial high-energy hot electron generation (~100 fs), enabling giant ultrafast nonlinearities. Crucially, at the communication band (1550 nm), $V_2C$ exhibits a high saturable absorption coefficient of -1.35 cm/GW. Integrating this into an erbium-doped fiber laser, we generate mode-locked pulses with a duration of 486 fs at 1569 nm, a 39.51 MHz repetition rate, and exceptional stability (92 dB SNR). This work establishes plasmonic MXenes as a paradigm for tailored ultrafast photonic devices.

**Introduction**

Plasmonic mode formed as collective oscillations of electrons lead to strong light-matter interactions, particularly in metallic nanoparticles, clusters, and structures, where the selected materials are typically metals with high electron concentrations[1]. When the free carrier concentration of the material is sufficiently high, plasmonic resonances can also be observed in semiconductors and oxides[2]. Nonequilibrium and high-energy charge electrons (above Fermi level) were generated in a plamonic nanostructure under sufficient excitation[3-4]. They exhibit pronounced strong filed location, high field enhancement and ultrafast hot electron evolution[5]. These characteristics perfectly meet the fundamental requirements of high-efficiency passively mode-locked lasers, enabling the generation of ultrashort pulses with low power consumption, high repetition rates, and ultrashort pulse durations.

Numerous studies have reported the combination of plasmonics with ultrashort pulse generation, with design strategies including the direct use of noble metals, the construction of composite systems containing noble metals, or the development of hybrid heterojunctions[6-8]. However, implementing this strategy within pure two-dimensional material systems has proven challenging, primarily because most two-dimensional materials exhibit semiconducting properties, making it difficult to generate surface plasmons during optical excitation. The emergence of 2D transition-metal carbides/carbonitrides (MXenes) material systems, which possess metallic properties, has made it possible to enhance the generation of ultrashort passive laser mode-locked pulses through the utilization of surface plasmon characteristics in a single two-dimensional material[9-10].

MXenes are a family of two-dimensional (2D) transition metal carbides, nitrides, and carbonitrides, first reported in 2011[11]. They are typically synthesized by etching A-layer elements from MAX phases ($M_{n+1}AX_n$), where M represents an early transition metal, A is a group 13 or 14 element, X is carbon and/or nitrogen, and n=1-3. MXenes possess a layered structure with metallic conductivity and abundant surface terminations (e.g., -OH, -F, -O). Their unique combination of high electrical conductivity, hydrophilicity, and mechanical flexibility makes them ideal for applications such as energy storage, electromagnetic shielding, water purification, catalysis, and biomedical devices[12]. They exhibit metallic properties that induce surface plasmon resonance, enabling strong and tunable plasmonic absorption in the visible and infrared regions. For instance, the carrier concentration of $Ti_3C_2$ (with electrons as the primary carriers) is approximately $10^{22}$ cm$^{-3}$ in a vacuum, indicating that plasmonic modes may exist in the visible to near-infrared region. The plasmonic mode at around 730 nm (1.7 eV) has been identified as a transverse plasmon mode[10]. Due to the symmetry of the particles and polarization breaking, the longitudinal mode is observed at much lower energies. Researchers have confirmed this through electron energy loss spectroscopy (EELS), where the longitudinal surface plasmon appears at 1350 nm (0.95 eV)[10]. Extending this model to other MXenes is crucial for understanding their optical properties. The free carrier concentration, which varies greatly depending on the MXene composition and structure, significantly affects its plasmonic modes[13]. The plasmonic absorption peak of $Ti_2C$ occurs at 542 nm, while $Ti_3C_2$ shows an absorption peak at 780 nm in the near-infrared region, corresponding to the transverse plasmon mode. This mode is the most prominent feature of most MXenes in their absorption spectra. Altering the transition metal elements can also influence this mode. The absorption peak of $Mo_2C$ shifts to 450 nm, while $Nb_2C$ shifts to 915 nm. $V_2C$ exhibits a noticeable absorption enhancement in the near-infrared region, which is characteristic of broad-spectrum plasmonic absorptions[14]. The nonlinear absorption response of $Ti_3C_2$[15], $Ti_3CN$[16], and $Nb_2C$[17] show typical nonlinear absorption mode conversion near their corresponding plasmonic characteristic wavelength. Such a conversion can be attributed to decrease in the photon energy is accompanied by a nonlinear optical response transition from saturable absorption to optical limiting. Due to unmatched surface plasmon response wavelength, these materials are clearly suboptimal for nonlinear optical modulation in the optical communication band. Therefore, the development of high-performance nonlinear materials, particularly for the near-infrared region and the communication C-band (1550 nm), is critically needed.

To address this challenge, we demonstrated surface plasmon induced enhanced ultrafast nonlinear saturable absorption response of $V_2C$ MXenes and achieved ultrashort pulse generation (~486 fs) in the communication C-band. Long-term stable few-layer $V_2C$ was prepared through an improved ion-exchange and organic solvent dispersion scheme. A theoretical model for the electron relaxation process of surface plasmon decay in few-layer $V_2C$ was proposed by combining transient absorption spectroscopy experiments and first-principles calculations. It was pointed out that the ultrafast response intensity of few-layer $V_2C$ is related to the generation of high-energy electrons near the surface of the nanostructure. By systematically comparing the nonlinear optical absorption responses of few-layer $V_2C$ under different excitation wavelengths, it was found that few-layer $V_2C$ can produce ultra-strong and ultrafast nonlinear optical responses at surface plasmon resonance wavelength 1550 nm. We achieved high-performance mode-locking in an erbium-doped fiber laser using our designed $V_2C$ MXene saturable absorber. The generated pulse had a full-width at half maximum (FWHM) of approximately 486 fs, a repetition rate of 39.51 MHz, and a signal-to-noise ratio (SNR) of 92 dB, centered at 1569 nm.

## Results

The proposed concept of plasmon-driven ultrashort pulse generation is illustrated in Figure 1. Figure 1a and b illustrate the schematic representation of the surface plasmon induced carriers relaxation processes, which involve several stages. (1) Photon absorption and plasmon excitation: The initial photon absorption induces collective oscillation of electrons, creating a surface plasmon mode that rapidly decays (~20 fs) through Landau damping, resulting in a non-equilibrium carrier distribution. (2) carrier distribution formation: Two primary mechanisms produce the carrier distribution. First, plasmon energy thermalizes electrons near the Fermi level, forming a distribution described by the Drude model. Alternatively, boundary constraints within the nanostructure can disrupt momentum conservation near the interface, breaking system translational symmetry and generating high-energy nonthermal hot electrons and hot holes, with energy levels reaching up to the energy of the pump photon. These high-energy carriers undergo energy redistribution through electron-electron scattering, typically lasting around a few hundred femtoseconds. (3) electron-phonon scattering and lattice heating: The thermalized carriers then exchange energy with lattice phonons, depleting additional electron energy and increasing lattice temperature over a timescale of several picoseconds. The rapid electron-electron scattering process, dominated by high-energy hot electrons, is crucial for all-optical devices, especially in applications like all-optical switching[18] and sub-picosecond pulse generation[19]. Hot electron yield scales inversely with plasmon photon energy squared, favoring low-energy excitation.[20] Consequently, by leveraging such a plasma-driven ultrafast hot electron relaxation mechanism within MXene's broad plasmon absorption band, ultrastrong and ultrafast nonlinear optical responses are anticipated.

The few-layer $V_2C$ nanosheets were prepared by combining ion exchange and solvent exchange, as described in Supplementary Note S1. As shown in Figure 2a, it can be seen that the multilayer $V_2C$ nanosheets exhibit a typical "accordion-like" structure, which is one of the distinctive evidence of the successful removal of the Al atomic layer in the MAX phase[17]. After further intercalation and ion exchange, few-layer $V_2C$ nanosheets were obtained. Figure 2b presents the transmission electron microscope (TEM) image of few-layer $V_2C$ nanosheets. The transparent flakes structure indicates that the prepared $V_2C$ nanosheets possess atomic-level thickness with only several atomic layers. The energy-dispersive X-ray spectroscopy (EDS) mapping results are shown in Figure S1. The four typical elements of V (yellow), C (green), F (red), and O (blue) overlap well, which also confirms the successful preparation of $V_2C$ nanosheets with surface -F and -O groups. As shown in Figure 2c, the XRD pattern of the $V_2AlC$ MAX phase ceramic closely matches the standard PDF card (JCDS PDF 29-0101) in peak position and intensity, confirming the high quality of the precursor. Compared to $V_2AlC$, the XRD characteristic peaks of few-layer $V_2C$ are either weakened or disappeared entirely, retaining only the (002) crystal plane alignment. Additionally, the (002) peak becomes broader and shifts to a lower angle at 7.4°, corresponding to an average interlayer spacing of 11.9 Å, which is significantly larger than the 6.6 Å interlayer spacing corresponding to the (002) peak of $V_2AlC$.

That verify the successful exfoliation of the MAX phase ceramic into a two-dimensional layered MXene. Figure 2d shows the Raman spectra of few-layer $V_2C$ nanosheets. The characteristic peaks at 181 cm$^{-1}$ and 267 cm$^{-1}$ are attributed to the in-plane $E_{2g}$ and $E_{1g}$ vibrational modes of $V_2C$ itself, while the peaks at 411 cm$^{-1}$, 526 cm$^{-1}$, and 657 cm$^{-1}$ arise from mixed in-plane/out-of-plane vibrational modes due to the interaction between different surface functional groups and the $V_2C$[21-22]. Figure 2e shows the absorption spectra of few-layer $V_2C$ dispersion at various concentration. The pronounced absorption enhancement in the near-infrared region is attributed to broadband plasmonic absorption. Additionally, in order to supplement the missing absorption spectrum in the near-infrared band （＞1500 nm）, elliptical polarisation spectrometer tests were carried out on the $V_2C$, and the results are shown in Figure S2. The variation of its extinction coefficient (k) proves that its strong plasmonic absorption peak can extend up to 2000 nm wavelength. The absorption spectra of few-layer $V_2C$ with different lateral size are shown in Figure S3. It can be seen that as the lateral size increases, the plasmonic absorption mode in the near-infrared band gradually weakens, similar to the broadening of the mode of gold nanoparticles[23-24]. Furthermore, the solvent stability of $V_2C$ was investigated by dispersing few-layer $V_2C$ in deionized water, N-Methylpyrrolidone (NMP), ethanol, N,N-dimethylformamide (DMF), and isopropanol (IPA), respectively. The temporal evolution of absorbance was then monitored by UV-Vis spectrophotometer. The resulting data, with time as the x-axis and normalized absorbance as the y-axis, are plotted in Figure 2f. The single exponential function was used to fit the data in the

Figure 2f[25-26], and the decay time constants of the different dispersants were 3949.8 h (NMP-$N_2$), 318.9 h (NMP), 72.3 h (DMF), 57.3 h (ethanol), 51.9 h (IPA), and 11.8 h (water), respectively. It can be concluded that effective isolation from water and oxygen in the NMP solvent and filled with $N_2$ is more conducive to the preservation of $V_2C$ nanosheets.

To investigate plasmon driven hot electron relaxation processes, we carried out transient absorption spectroscopy measurements of few-layer $V_2C$. In conventional transient absorption spectroscopy measurements of semiconductors, both the pump and probe photon energies exceed the semiconductor bandgap[27]. The sample absorbs the pump light, inducing interband transitions that create population in excited states. This population alters the dielectric function of the sample, thereby modifying the absorption of the probe light. Similarly, Drude electrons and hot electrons generated through plasmonic mode absorption also affect carrier distributions, leading to changes in the sample's dielectric function and producing transient absorption signals[28]. Figures 3(a) and (b) present transient absorption spectra of few-layer $V_2C$ at pump wavelengths of 325 nm and 1300 nm, corresponding to interband transition and plasmonic excitation, respectively. Under both pumping modes, the few-layer $V_2C$ nanosheets exhibits broadband photo-induced absorption (PIA) signals (positive ΔA) at probe wavelengths below 670 nm. This phenomenon is conventionally ascribed to the excited-state absorption (ESA) process of the probe light, where the electrons in excited states absorb the probe light to reach higher excited states, resulting in an increase in the absorption of the probe light. Interestingly, the signal with a probe wavelength greater than 670 nm significantly changes from a positive signal to a negative signal, corresponding to the photobleaching (PB) process. That is, the electrons in the excited state block the absorption of the probe light, resulting in a weakened absorption of the probe light.

Figures 3(c-f) show dynamics for two distinct pump wavelengths and probe wavelengths. At a probe wavelength of 1100 nm (Figures 3d and 3f), an ultrafast component close to the pulse width appears immediately after excitation, followed by a typical picosecond-scale electron-phonon coupling relaxation. This ultrafast component is attributed to high-energy hot electrons generated through electron-electron scattering in the plasmonic mode, as previously described. At a probe wavelength of 500 nm (Figures 3c and e), the ultrafast component is diminished, and electron-phonon scattering on the picosecond scale predominates, underscoring the plasmon mode's significance in generating high-energy surface charges. For the 325 nm pump and 500 nm probe wavelengths (Figure 3c), the ultrafast component almost disappear which indicates that interband absorption does not produce hot electrons in this condition.

The dynamics were fitted using a bi-exponential decay model, where the first fast component $\tau_1$ and its corresponding amplitude $A_1$ vary with the probe wavelength, as shown in Figures 3g and h, respectively. As anticipated, when the pump light is at 1300 nm with probe wavelengths exceeding 670 nm, the time corresponding to $\tau_1$ is the shortest, reaching around 100 fs. Moreover, the proportion corresponding to A1 is the largest, reaching 0.9. This indicates that the electron-electron scattering process induced by high-energy hot electrons dominates at this point, accounting for up to 90% of the entire relaxation process. When the wavelength of probe light is less than 670 nm, $\tau_1$ increases to approximately 1 ps under both pump modes, and the proportion of the fast response decreases to around 40%. This indicates that the relaxation process is dominated by electron-phonon scattering on the picosecond timescale (~5 ps). This agrees with Landau-Fermi liquid theory[29], where high-energy electron relaxation involves effective electron-electron scattering, while low-energy excitations near the Fermi sea are governed by both electron-electron and electron-phonon scattering. Thus, photobleaching signals at probe wavelengths above 670 nm stem from high-energy carrier relaxation in the plasmonic mode, while photobleaching at shorter probe wavelengths relates to low-energy relaxation from interband transitions. Figure 3i shows the transient spectral signals of $V_2C$ under different pump wavelength excitation with a delay time of 400 fs. It can be seen that when the pump power is nearly the same, the amplitude value of the PIA signal corresponding to the pump wavelength of 1300 nm is almost five times that of the pump wavelength of 325 nm. Moreover, when the pump wavelength is 325 nm, the PB signal in the near-infrared region is weaker. In addition, by comparing the ratios of PIA and PB signals at the two pump wavelengths, it was found that the ratio at a pump wavelength of 325 nm was much higher than that at 1300 nm. This indicates that the proportion of hot carriers generated at a pump wavelength is near-infrared band (1300 nm pump) is much higher than that of visible band (325 nm).

First-principles calculations further analyzed the excited-state dynamics of $V_2C$,

with computed band structures. Vertical transitions require photon energy matching the bandgap as well as opposite parity between the initial and final state wavefunctions. The transition intensity depends on the joint density of states (JDOS), with van Hove singularities at high-symmetry points contributing significantly[30]. Considering the different surface functional groups of $V_2C$, the band structures and density of states (DOS) for $V_2CO_2$ and $V_2CF_2$ were calculated, as shown in Figures 4a and 4b. Consistent with previous studies[21, 31-32], both $V_2CO_2$ and $V_2CF_2$ exhibit metallic properties due to the crossing of their electronic bands with the Fermi level. Due to the limitations of first-principles calculations, high-energy electronic excitations are not considered here. Instead, the focus is on low-energy electrons near the Fermi level, which are typically generated by Drude electrons and interband transitions induced by plasmonic excitations[33]. When analyzing the symmetry of band structures relevant to excited-state absorption, the bands near the Fermi level are regarded as initial states, while those above the Fermi level serve as final states. Moreover, the van Hove singularities near the Γ-point, characterized by the shape of the bands, are most likely to facilitate optical vertical transitions. Therefore, only the symmetry and transitions at the Γ-point are considered in this study. For instance, in $V_2CO_2$, the parities near the Fermi level are even, suggesting that the wavefunctions of excited-state electrons are of odd parity. Conversely $V_2CF_2$, the parities near the Fermi level are odd, indicating that the excited-state electron wavefunctions exhibit even parity. Consequently, for the excited-state transitions of $V_2CO_2$ and $V_2CF_2$, electrons must transition to bands with even and odd parity, respectively. As depicted by the green and red arrows in the figure, the excited-state absorption transitions near the Fermi level require relatively high photon energies (wavelengths below 600 nm). The energy positions of the final states for these transitions correspond to peaks in the density of states, potentially leading to significantly enhanced optical transitions. This correlates with the experimentally observed enhancement of probe light absorption (PIA). On the other hand, the photon energy indicated by the pink arrow corresponds to 1100 nm, where no parity-allowed electronic bands are available near the terminal states. Consequently, the excited-state absorption transition is prohibited, and it is more inclined to generate the PB signal caused by the plasmon mode, which is consistent with the observed results in the experiment (the probe light absorption is weakened). Figure 4c shows the JDOS distribution for $V_2CO_2$ and $V_2CF_2$, representing the photon energy-dependent transition intensity from ground states (below the Fermi level). A peak in the JDOS is observed at approximately 4 eV, indicating the strongest interband transition intensity, which matches the 325 nm pump wavelength used in our transient absorption experiments. Combining experimental results with first-principles calculations, a schematic of the $V_2C$ optical absorption transitions is proposed in Figure 4d. For photon energies below 4 eV, the optical absorption process is dominated by plasmonic modes. Subsequently, part of the plasmon energy is transferred to thermalize electrons near the Fermi level, forming a distribution described by the classical Drude model. Another portion of the plasmon energy generates high-energy, non-thermalized hot electrons and hot holes via momentum-nonconserving intraband transitions near the nanostructure surface.

The maximum energy of these hot carriers can reach the photon energy of the pump light.

Moreover, we performed nonlinear optical absorption measurements on few-layer $V_2C$ in the near-infrared region to verify its plasmon-driven broadband giant ultrafast nonlinear optical response. We selected excitation wavelengths of 800 nm, 1150 nm, 1300 nm, 1550 nm, and 1800 nm for the Z-scan experiments. Figures 5a-e show the normalized transmittance of the few-layer $V_2C$ sample versus its relative position to the lens focal point along the Z-axis during the open-aperture Z-scan tests at different excitation wavelengths. At all excitation wavelengths, they exhibited the nonlinear optical saturation absorption response, represented by symmetric peak shapes around the focus. This suggests that when excited by high-intensity near-infrared laser light, few-layer $V_2C$ nanosheets show a decrease in absorption, similar to the optical bleaching response observed in transient absorption experiments, confirming the consistency of the two experimental results. Moreover, as the excitation light intensity increased, the experimental peak signal showed a linear increase, indicating that the Z-scan signal was caused by the intrinsic optical nonlinearity of the few-layer $V_2C$, rather than other scattering effects. To further quantify and compare the optical nonlinearity at different wavelengths, the following nonlinear optical theory is used to fit the experimental data and obtain results that can be used for quantitative comparison.

The open-aperture Z-scan experimental curves can be fitted by[34]:

$$T = \frac{1}{\sqrt{\pi}q_0} \int_{-\infty}^{\infty} \ln\left[1 + q_0 \exp(-x^2)\right] dx$$

where $q_0 = \alpha_{NL}I_0L_{eff}$, $\alpha_{NL}$ is the nonlinear absorption coefficient, $I_0$ is the on-axis peak intensity on the focal plane, $L_{eff} = (1-e^{\alpha_0 L})/\alpha_0$ is the sample's effective thickness, where $\alpha_0$ is the linear absorption coefficient and L is the thickness of the sample. By fitting the Z-scan curve, the $\alpha_{NL}$ can be obtained for different cases. As shown in Figure 5f, the $\alpha_{NL}$ of few-layer $V_2C$ at 800 nm, 1150 nm, 1300 nm, 1550 nm, and 1800 nm were determined to be −0.05 cm/GW, −0.72 cm/GW, −0.80 cm/GW, −1.35 cm/GW, and −0.46 cm/GW, respectively. Notably, at an excitation wavelength of 1550 nm, the absolute value of the absorption coefficient is largest, indicating that few-layer $V_2C$ exhibits the best saturation absorption performance at this wavelength. In comparison to $Nb_2C$, which also demonstrates excellent nonlinear absorption properties in the near-infrared band, few-layer $V_2C$ at 1550 nm has a nonlinear absorption coefficient of −1.35 cm/GW, which is double that of $Nb_2C$ at 1200 nm (−0.60 cm/GW)[17], highlighting the unmatched advantage of few-layer $V_2C$ as a nonlinear optical device in the near-infrared region.

The proof-of-concept application of ultrafast optical nonlinearity in few-layer $V_2C$ was demonstrated by realizing mode locking in an erbium-doped fiber laser. The laser mode-locking setup follows previous works[35-36]. Briefly, a 976 nm laser diode was used as the pump source, with a custom hybrid cavity comprising an integrated wavelength division multiplexer, output coupler, and polarization-independent isolator. The cavity was coupled to a side-polished optical fiber to create an all-fiber saturable absorber. In the experiment, when the pump power was increased to 72 mW, stable mode-locking operation was achieved by carefully adjusting the polarization inside the cavity. The mode-locking state was maintained until the pump power reached a maximum of 500 mW, with an output power of 12.24 mW. Figure 6a displays the pulse autocorrelation trace, where the single pulse profile was fitted using a Sech² function, yielding a full-width at half maximum (FWHM) of approximately 486 fs. Figure 6b records the same pulse sequence over 1000 ns, showing that the intensity of each pulse remains nearly constant, with a pulse interval of 253 ns. In addition, Figure 6c shows the radio-frequency spectrum of the fiber laser, where the mode-locked pulse repetition rate is 39.51 MHz, with a signal-to-noise ratio (SNR) of 92 dB, indicating high stability of the mode-locking operation. Figure 6d shows the output spectrum after successful mode-locking, with distinct Kelly sidebands, confirming that the fiber laser operated in soliton mode. The 3 dB bandwidth and central wavelength of the output spectrum were 5.9 nm and 1569 nm, respectively. Furthermore, the long-term stability of the mode-locked fiber laser system was verified by recording the output spectrum every hour for 8 hours, as shown in Figure 6d. No collapse or significant changes in the output spectrum were observed, confirming the high stability of the mode-locked fiber laser based on few-layer $V_2C$. In contrast, previously reported typical MXenes near 1550 nm had mode-locked output pulse durations of 660 fs ($Ti_3CN$)[37], 850 fs ($Ti_3C_2$)[38], 603 fs ($Nb_2C$)[39], and 1.81 ps ($Mo_2C$)[40], all of which have much longer pulse durations than the 486 fs pulse duration obtained in this work, further demonstrating the potential of $V_2C$ as an ultrafast photonics device at this wavelength.

**Discussion**

In summary, we establish a robust synthesis protocol for few-layer $V_2C$ MXene via optimized ion-exchange and organic-solvent dispersion, achieving unprecedented long-term stability. Critically, we uncover a new electron relaxation mechanism driven by surface plasmon decay in $V_2C$ MXene, modeled through a first-principles-informed theoretical framework and experimentally validated by transient absorption spectroscopy. This reveals that high-energy hot electron generation at the nanostructure interface dictates the material's ultrafast nonlinear intensity. Leveraging this fundamental insight, we demonstrate that 1550 nm excitation triggers ultrastrong and ultrafast nonlinearities in $V_2C$. Specifically, we engineered an ultrafast fiber laser directly harnessing this nonlinear response, achieving mode-locked pulses at 486 fs near 1550 nm. This represents the implementation of plasmon-driven MXene-based nonlinear control in femtosecond fiber lasers, establishing a new paradigm for tailored ultrafast photonics.

## Materials and methods

## Preparation of Solution-Stable $V_2C$

We combined ion and solvent exchange methods to achieve a solution-stable few-layer $V_2C$ that can be preserved long-term. The preparation steps are as follows: First, prepare 50 ml of 49% HF and pour it into a Teflon reactor, then place the reactor in a temperature-controlled shaker (40°C, 500–800 rpm) for half an hour. Gradually add 2 g of $V_2AlC$ precursor powder in small portions to prevent excessive reaction intensity. After 48 hours, remove the reactor and centrifuge the contents, carefully transferring the solution to several 50 ml centrifuge tubes. Set the centrifuge to 5000 rpm for 5 minutes. Discard the supernatant, add fresh deionized water to the tubes, shake well, and repeat centrifugation at the same settings until the supernatant reaches a pH of 6. Add 25 ml of 5% TMAOH intercalation agent to the centrifuged precipitate and stir at room temperature for 12 hours. Afterward, remove the excess intercalation agent by centrifuging at 6500 rpm for 5 minutes, discarding the supernatant, and adding ethanol to the precipitate. Shake well for 10 minutes, centrifuge again, and repeat three times, measuring the pH each time until it is close to 7. Add half the amount of deionized water to the precipitate and ultrasonicate in an ice bath for 1 hour. Then centrifuge at 3000 rpm for 15 minutes. During centrifugation, prepare a saturated LiCl solution (10 g LiCl in 100 ml of water). After centrifugation, add the supernatant to the saturated LiCl solution, stir in a beaker, let it stand for 12 hours, and then decant the supernatant. Add another 100 ml of saturated LiCl solution to the precipitate and let it sit for 12 hours. The resulting flocculates can be preserved long-term and used to obtain few-layer $V_2C$ on demand through washing. Using a dropper, transfer the flocculates from the previous step into a centrifuge tube, centrifuge at 3500 rpm for 5 minutes, discard the supernatant, and add deionized water to the precipitate, shaking thoroughly. Repeat this process three times. After the final centrifugation, add a certain amount of N-methyl-2-pyrrolidone (NMP) to the precipitate, shake well, and ultrasonicate for 30 minutes. Centrifuge at 3000 rpm for 15 minutes, collecting the upper liquid as few-layer $V_2C$ dispersed in NMP. This dispersion can be preserved for extended periods under nitrogen gas.

**Morphological and spectral characterization**

Transmission electron microscopy (TEM) was conducted on a JEM-3100 instrument (JEOL, Japan). X-ray diffraction (XRD) measurements were carried out with a Bruker D8 Advance diffractometer. Raman spectroscopy was performed on a confocal Raman system (inVia Qontor, Renishaw, UK). UV–vis–NIR absorption spectra were obtained using a Cary60 UV–vis spectrophotometer (Agilent).

**Transient absorption spectrum and OA Z-scan**

The light source was a mode-locked Ti:sapphire regenerative amplifier system (Astrella-Tunable-USP-1K, Coherent Inc.) operating at 800 nm, a 35 fs pulse, with a repetition rate of 1 kHz. The wavelength tuning was achieved using two optical parametric amplifiers (TOPAS-Prime, Coherent Asia,Inc.), which subsequently served as the light source for both transient absorption and Z-scan measurements. Transient absorption spectra were acquired using a Helios transient absorption spectrometer (Helios, Ultrafast Systems Inc.). The nonlinear optical properties were characterized by a custom-built Z-scan system.

### DFT calculation.

The VASP code was employed for density functional theory[41] calculations with the projector-augmented wave method[42]. Since the strong correlation effects of V atoms, the generalized gradient approximation (GGA) plus Hubbard U[43] method was employed during the calculations, the Ueff value of V 3d orbital was set to be 4 eV[44]. The plane-wave energy cutoff was set to 500 eV, and k-point sampling 9x9x1 grid with Monkhorst-Pack scheme. The initial structure symmetry (hexagonal, $p\bar{3}m1$ space group) of $V_2CF_2$ and $V_2C(OH)_2$ were kept after fully relaxed until the force between each atom is less than 0.01 eV/Å. The band symmetry and parity analyses were carried out in QUANTUM ESPRESSO package[45].

## Acknowlegement

This research was supported by the National Natural Science Foundation of China (Nos. 62275275, 11904239, 62422506, 12474383, 52273202), National Key R&D Program of China (2022YFA1604200), Natural Science Foundation of Hunan Province (Grant Nos. 2021JJ40709, 2022JJ20080, 2024JJ6481) and Postgraduate Innovative Project of Central South University (Grant No. CX20230246). This work was also supported in part by the High-Performance Computing Center of Central South University and Open Sharing Found for the Large-scale Instruments and Equipment of Central South University. We also acknowledge resources from the Hefei Advanced Computing Center.

## Author Contributions

Y. W. Wang and J. He conducted and supervised the project. Y.D.Wang fabricated the samples, conducted a series of data collection and analysis. Y. D. Wang carried out the density functional theory calculations. L. Zhou, Q. Long, H. Chang and S. Xiao helped with the optical tests. N. Li, C. Zhao and S. Wen carried out ultrashort pluses generation. All authors contributed to the redaction of the manuscript and agreed with the final version of the manuscript.

## Data availability

The data that support the findings of this study may be obtained from the corresponding authors upon request.

## Conflict of interests

The authors declare no conflicts of interest.

# Figure legends

**Figure 1. Schematic illustration of plasmon-driven ultrashort pulse generation.** (**a**). Schematic of surface plasmons generation on 2D $V_2C$. (**b**). The surface plasmon induced carriers relaxation processes. (**c**) and (**d**). Schematic diagram of erbium-doped fiber laser and output ultrashort pulse.

**Figure 2. Structure and stability characterization of the $V_2C$ nanosheets.** (**a**) SEM image of multilayers $V_2C$. (**b**) TEM image of few-layer $V_2C$ nanosheets. (**c**) XRD pattern of $V_2C$ (top), $V_2AlC$ (middle), and JCDS PDF#29-0101 ($V_2AlC$) (bottom). (**d**) Raman spectra of $V_2C$ nanosheets. (**e**) Absorbance of $V_2C$ solution with different concentrations. (**f**) The absorbance of few-layer $V_2C$ varying with time in different solvents.

**Figure 3. Plasmon induced carriers relaxation processes measurements.** Transient absorption spectra of $V_2C$ with pump wavelengths of (**a**) 325 nm and (**b**) 1300 nm. (**c-f**). Temporal Dynamics dynamic processes at different pump/probe wavelengths. Fitted time constant (**g**) and amplitude (**h**) of the first fast process at different pump/probe wavelengths. (**i**). The transient spectral signals of $V_2C$ at a delay of 400 fs under different pump wavelength excitations.

**Figure 4. First-principles calculation analysis of plasmon induced carrier.** (**a**) Band Structure and Density of States of $V_2CO_2$ and (**b**) $V_2CF_2$. (**c**) Joint density of states distribution of $V_2CO_2/V_2CF_2$. (**d**) Illustration of $V_2C$ Electronic Transitions

**Figure 5. Broadband nonlinear saturable absorption response of $V_2C$.** (**a-e**) Z-scan experimental curves and theoretical fitting results of few-layer $V_2C$ at different excitation wavelengths. (**f**) Nonlinear optical absorption coefficient as a function of wavelength.

**Figure 6. Plasmon-driven highly efficient ultrashort pulse generation.** (**a**) pulse duration, (**b**) oscilloscope signal, (**c**) radio-frequency spectrum, (**d**) output spectrum and long-term stability.

**Figure 1.**

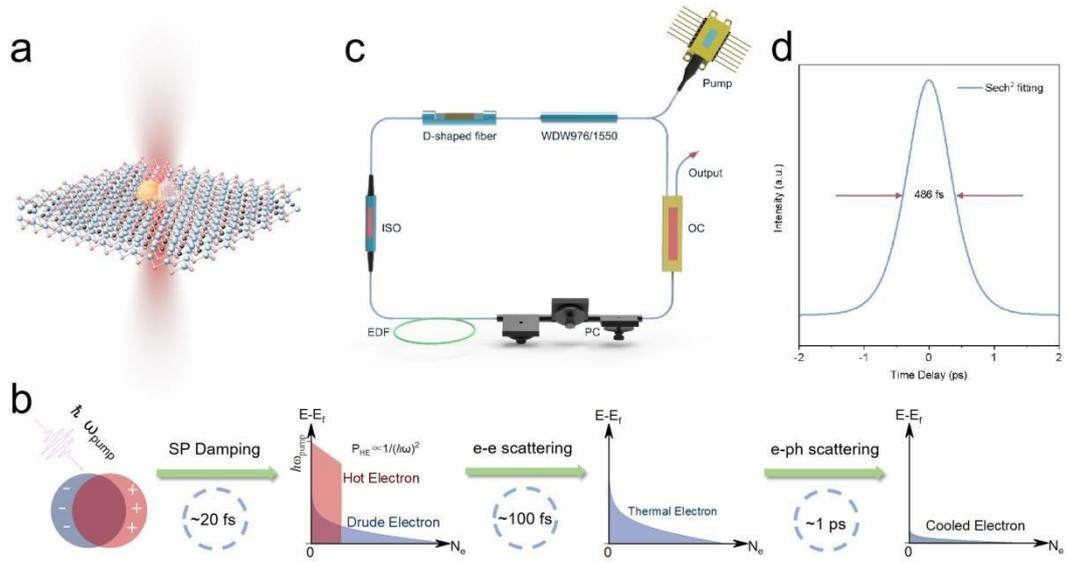

**Figure 2.**

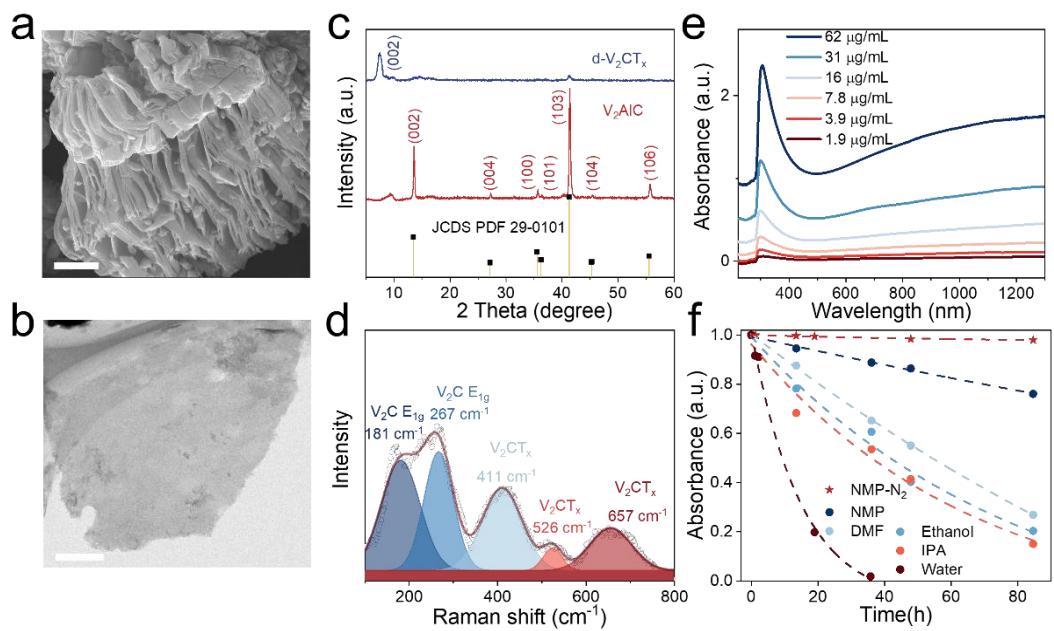

**Figure 3.**

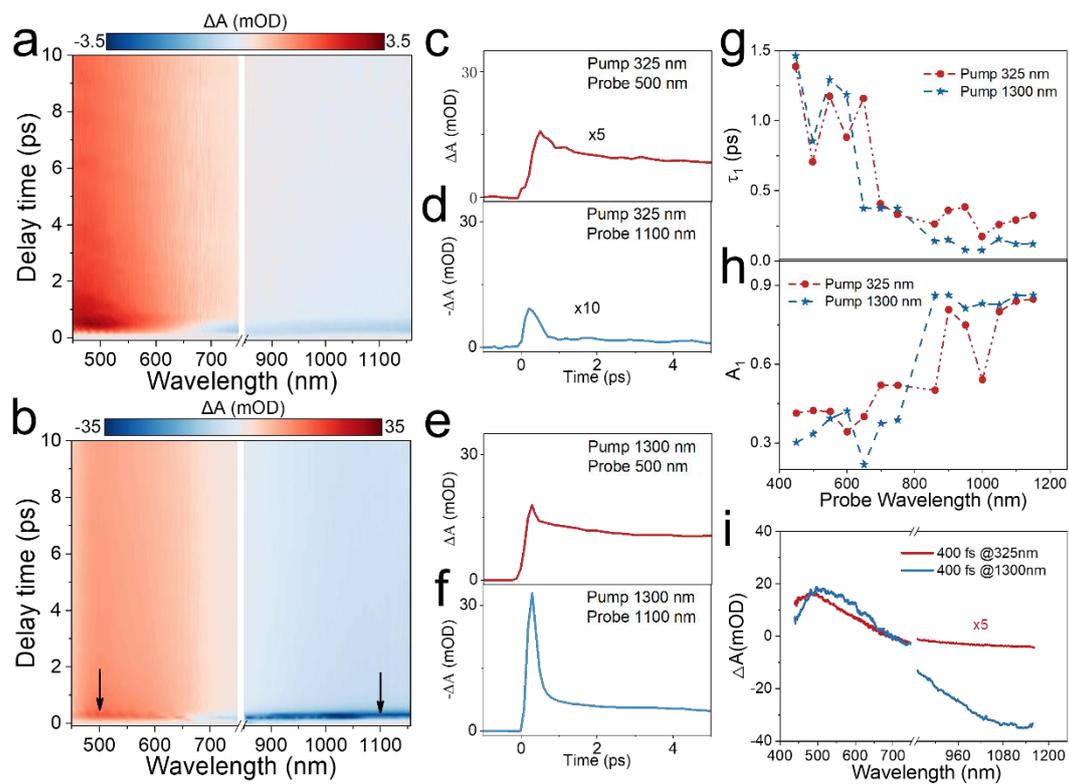

**Figure 4.**

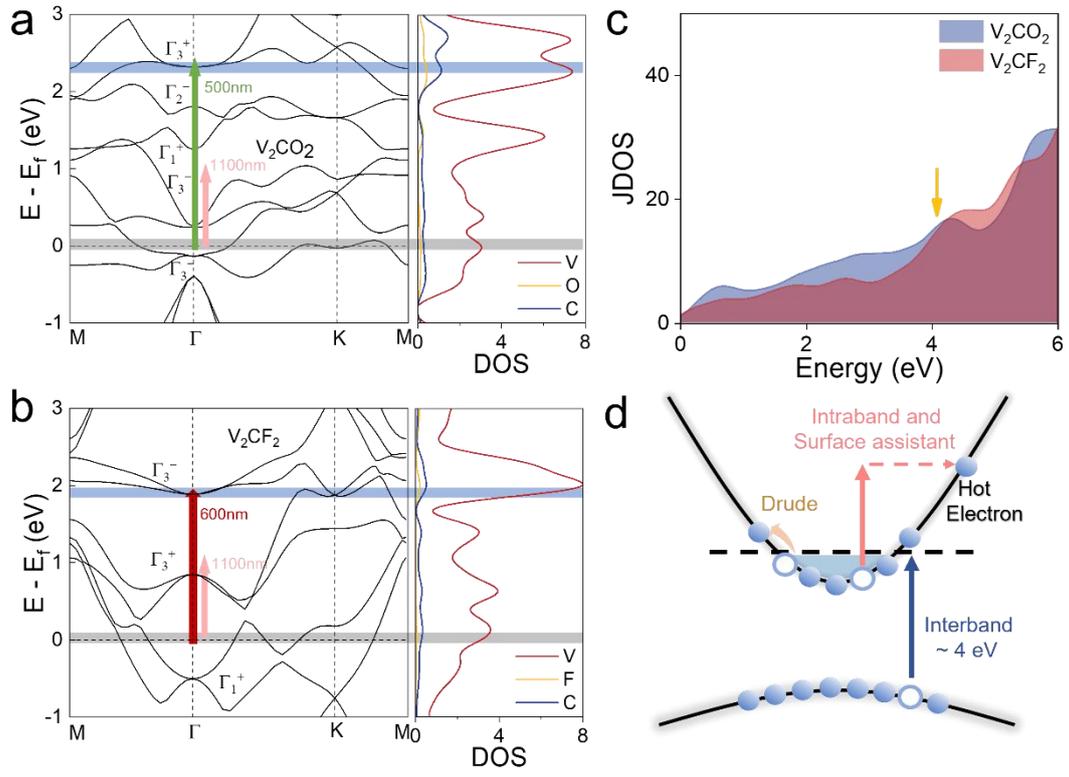

**Figure 5.**

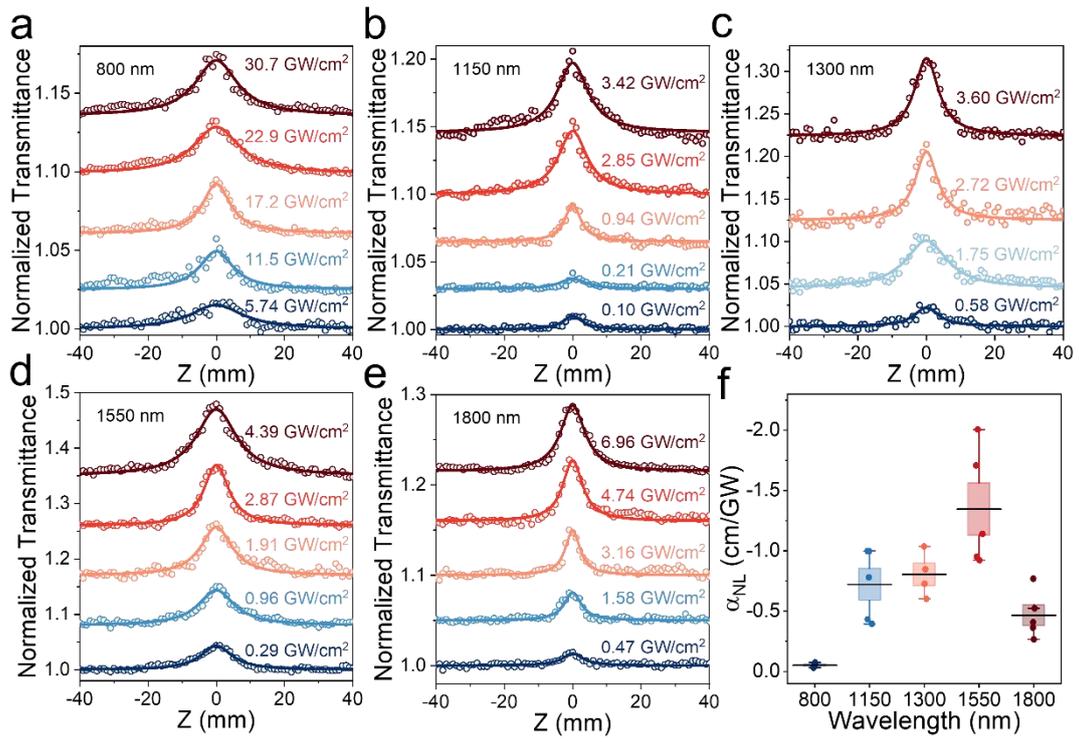

**Figure 6.**

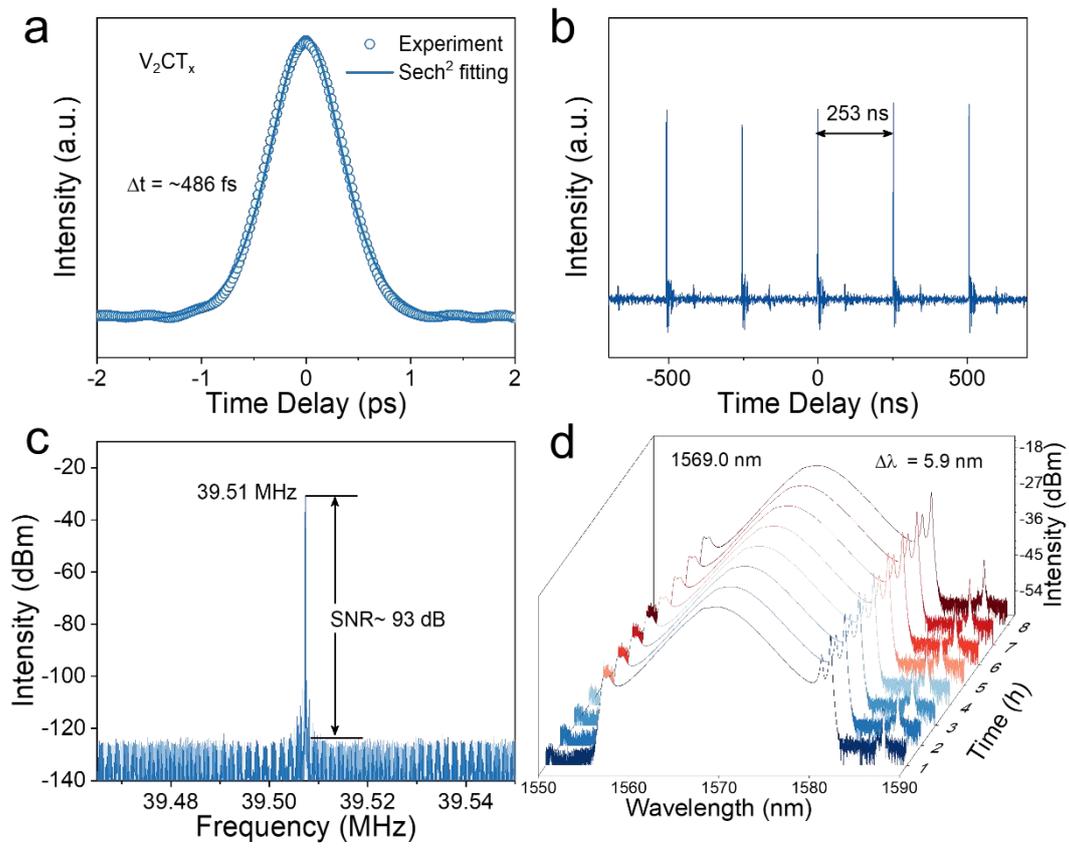

# Supplementary Materials for

# Plasmon-driven Ultrafast and Highly Efficient Saturable Absorption for Ultrashort Pulse Generation Based on 2D V$_2$C

Yingwei Wang[1,*], Li Zhou, Quan Long[1], Xin Li[1,4], Haolin Chang[1], Ning Li[3], Yiduo Wang[1,2,*], Bei Zhang[4], Zhihui Chen[1], Zhongjian Yang[1], Si Xiao[1], Chujun Zhao[3,*], Shuangchun Wen[3], Jun He[1,*]

[1]Hunan Key Laboratory of Nanophotonics and Devices, School of Physics, Central South University, Changsha 410083, China

[2]Beijing National Laboratory for Condensed Matter Physics, Institute of Physics, Chinese Academy of Sciences, Beijing 100190, China

[3]Key Laboratory for Micro- and Nano-Optoelectronic Devices of Ministry of Education, College of Physics and Microelectronic Science, Hunan University, Changsha 410082, China

[4]School of Physics Science and Technology, Xinjiang University, Urumqi, Xinjiang 830046, China

* Email: wyw1988@csu.edu.cn; yiduowang@iphy.ac.cn; cjzhao@hnu.edu.cn; junhe@csu.edu.cn;

**Supplementary Note S1: Preparation of Solution-Stable V$_2$C**

MXenes with M$_2$X structures are generally more chemically active than those with M$_3$X$_2$ structures, due to their higher active surface area per unit mass (e.g., V$_2$C has three atomic layers compared to Ti$_3$C2's five atomic layers). Among all M$_2$X structures, vanadium-based MXenes are particularly reactive, making V$_2$C one of the least stable MXenes. Single-layer and few-layer V$_2$C degrade readily when dispersed in water or exposed to air. This instability of few-layer V$_2$C in water primarily results from large organic cations (such as TMA$^+$ or TBA$^+$) introduced during intercalation[1]. A recent approach suggests using Li$^+$ ions to replace these large cations, generating flocculates that can be preserved for over three months, allowing the preparation of few-layer V$_2$C dispersions in water through subsequent washing[1]. However, even with ion exchange and flocculation, few-layer V$_2$C remains challenging to store in water over long periods. Here, we combined ion and solvent exchange methods to achieve a solution-stable few-layer V$_2$C that can be preserved long-term.

The preparation steps are as follows: (1) First, prepare 50 ml of 49% HF and pour it into a Teflon reactor, then place the reactor in a temperature-controlled shaker (40°C, 500–800 rpm) for half an hour. Gradually add 2 g of V$_2$AlC precursor powder in small portions to prevent excessive reaction intensity. (2) After 48 hours, remove the reactor and centrifuge the contents, carefully transferring the solution to several 50 ml centrifuge tubes. Set the centrifuge to 5000 rpm for 5 minutes. Discard the supernatant, add fresh deionized water to the tubes, shake well, and repeat centrifugation at the same settings until the supernatant reaches a pH of 6 or higher. (3) Prepare 25 ml of 5% TMAOH intercalation agent, add it to the centrifuged precipitate, shake thoroughly (the solution should become viscous), and stir at room temperature for 12 hours. Afterward, remove the excess intercalation agent by centrifuging at 6500 rpm for 5 minutes, discarding the supernatant, and adding ethanol to the precipitate. Shake well for 10 minutes, centrifuge again, and repeat three times, measuring the pH each time until it is close to 7. Ethanol, instead of deionized water, is used here because MXenes do not disperse well in ethanol, but organic cations dissolve in it, making this step more efficient and thorough in removing organic ions. (4) Add half the amount of deionized water to the precipitate and ultrasonicate in an ice bath for 1 hour. Then centrifuge at 3000 rpm for 15 minutes. During centrifugation, prepare a saturated LiCl solution (10 g LiCl in 100 ml of water). After centrifugation, add the supernatant to the saturated LiCl solution, stir in a beaker, let it stand for 12 hours, and then decant the supernatant. Add another 100 ml of saturated LiCl solution to the precipitate and let it sit for 12 hours. The resulting flocculates can be preserved long-term and used to obtain few-layer V$_2$C on demand through washing. (5) Using a dropper, transfer the flocculates from the previous step into a centrifuge tube, centrifuge at 3500 rpm for 5 minutes, discard the supernatant, and add deionized water to the precipitate, shaking thoroughly. Repeat this process three times. After the final centrifugation, add a certain amount of N-methyl-2-pyrrolidone (NMP) to the precipitate, shake well, and ultrasonicate for 30 minutes. Centrifuge at 3000 rpm for 15 minutes, collecting the upper liquid as few-layer V$_2$C dispersed in NMP. This dispersion can be preserved for extended periods under nitrogen gas.

**Figure S1**

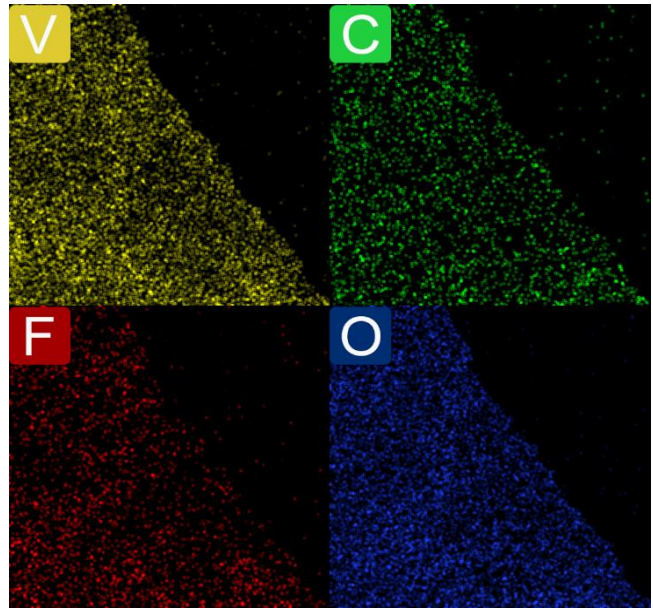

**Figure S1**. EDS mapping images of few-layer V$_2$C nanosheets.

**Figure S2**

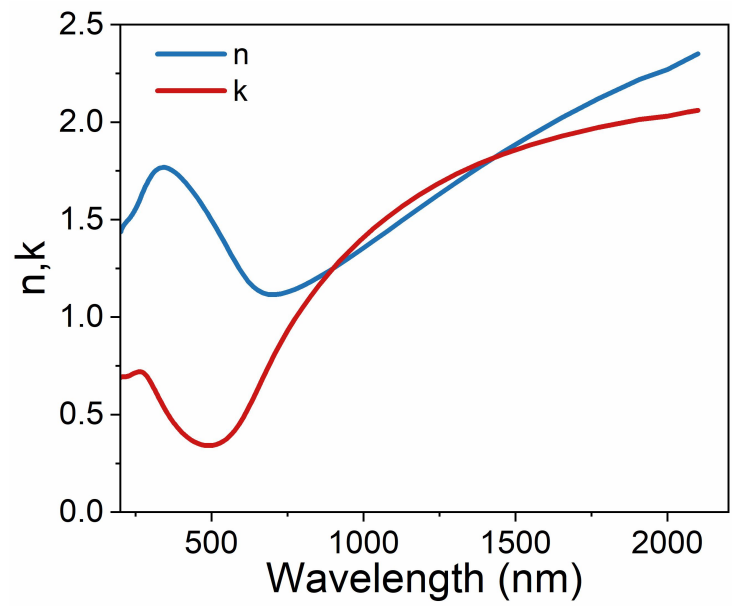

**Figure S2**. The fitting results of $V_2C$ elliptic polarization spectrometer test.

**Figure S3**

Previous studies on metal nanoparticles show that the plasmon absorption bandwidth increases as particle size decreases due to a reduced mean free path of free electrons in smaller particles[2, 3]. For gold particles smaller than 2 nm, the plasmon band broadens and eventually vanishes. Studies on $Ti_3C_2$ plasmon modes[4] indicate that the transverse plasmon mode is size- and shape-insensitive, whereas the longitudinal plasmon mode is highly sensitive to both. Similar behavior is expected for few-layer $V_2C$. By examining the absorption of few-layer $V_2C$ of varying lateral sizes, insights into the origin of its plasmonic absorption modes can be gained. Few-layer $V_2C$ dispersions were sonicated for 2 hours, then subjected to gradient centrifugation to obtain samples with different lateral sizes.

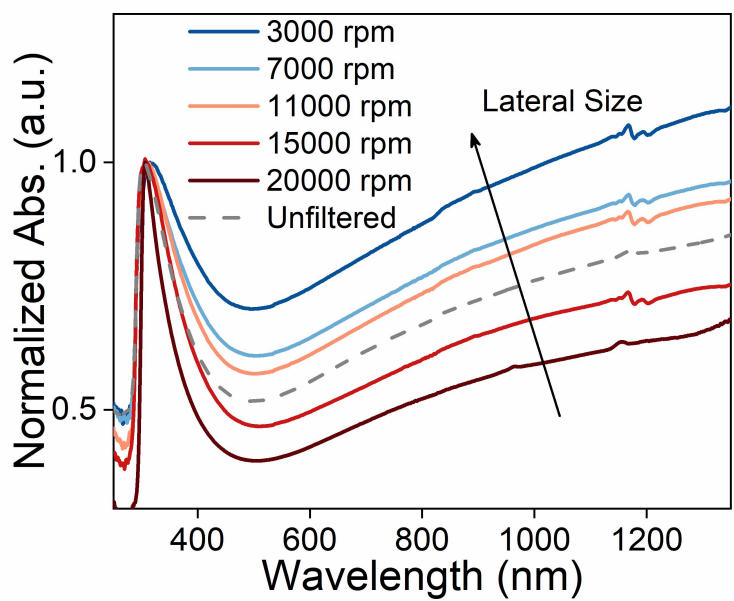

**Figure S3**. The absorption spectra of $V_2C$ with different transverse dimensions obtained under gradient centrifugation.